\documentclass[twocolumn,nobibnotes,prx]{revtex4-1}

\usepackage{graphicx}
\usepackage{amsmath}
\usepackage{tikz}
\usepackage{siunitx}
\usepackage{amsfonts}
\usepackage{natbib}
\usepackage[caption=false]{subfig}

\begin{document}
\title{A directional wave measurement attack against the Kish key distribution system}
\author{Lachlan J.~Gunn}
\email{lachlan.gunn@adelaide.edu.au}
\author{Andrew Allison}
\email{andrew.allison@adelaide.edu.au}
\author{Derek Abbott}
\email{derek.abbott@adelaide.edu.au}
\affiliation{School of Electrical and Electronic Engineering, The University of Adelaide, SA 5005, Australia}
\begin{abstract}
	The Kish key distribution system has been proposed
	as a classical alternative to quantum key distribution.  The idealized Kish
	scheme elegantly promises secure key distribution by exploiting thermal noise
	in a transmission line.  However, we demonstrate that it is
	vulnerable to nonidealities in its components, such as the finite
	resistance of the transmission line connecting its endpoints.  We introduce
	a novel attack against this nonideality using directional wave measurements,
	and experimentally demonstrate its efficacy.
    Our attack is based on causality: in a spatially distributed system,
    propagation is needed for thermodynamic equilibration, and that leaks information.
\end{abstract}
\maketitle


The Kish key distribution (KKD) system, based on Kirchhoff's laws
and Johnson noise (KLJN)~\cite{kish-kljn-original}
has been proposed as a classical alternative to quantum key
distribution (QKD)~\cite{bb84}.  Eschewing expensive and
environmentally-sensitive optics, it can be implemented economically
in a wider variety of systems than QKD.

\begin{figure}[h]
	\centering
	\includegraphics[width=0.65\linewidth]{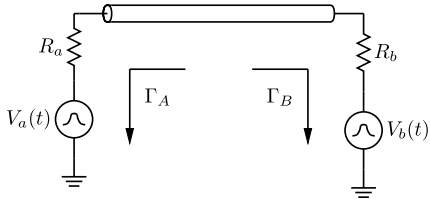}
	\caption{We determine the forward- and reverse-travelling waves in this
			idealized KKD system.  Practical systems would include
			low-pass filters and instrumentation that do not affect the
			steady-state signal.  The mean-squared voltages
			$\langle V_a^2(t) \rangle$ and $\langle V_b^2(t) \rangle$
			are proportional to the resistances $R_a$ and $R_b$
			respectively.
			We perform our analysis in terms of the reflection
			coefficients $\Gamma_A$ and $\Gamma_B$.}
	\label{fig:kljn-sparam}
\end{figure}

The KKD system is claimed~\cite{kish-kljn-original} to derive
unconditional security from the second
law of thermodynamics---the idea being that net power cannot flow from
one resistor to the other under equilibrium.

An idealised KKD system is shown in Figure~\ref{fig:kljn-sparam}.  Alice
and Bob each apply a noise signal to a line through a series resistor.
The voltage on the line is unchanged if the terminals of
Alice and Bob are swapped;
if the mean-square voltages applied by Alice and Bob are proportional
to $R_a$ and $R_b$ respectively then no average power flows 
through the line, and in the ideal case an eavesdropper, Eve,
cannot determine which end has which resistance~\cite{kish-kljn-original,gingl-kljn-analysis}.
If Alice and Bob randomly choose their resistances---resulting in
corresponding noise amplitudes---to be either $R_h$ or $R_l$, three possibilities
avail themselves: both choose $R_h$, both choose $R_l$, or
one chooses $R_h$ and the other chooses $R_l$.
In this third case, Alice knows the value of her own
resistor, and so can deduce Bob's resistor via noise spectral
analysis, and vice-versa.
However, an eavesdropper
lacks this knowledge, and so in the ideal case Alice and Bob
have secretly shared one bit of information.

It has been claimed~\cite{kish-notes-on-recent-approaches} that
transmission line theory does not apply to the
the KKD system when operated at frequencies below
$f_c = \nu/({2L})$,
where $L$ is the transmission line length and $\nu$ the signal propagation
velocity, because wave modes do not propagate
below this cutoff.  We demonstrate that
this is not the case by constructing a directional wave measurement device that
is then used for a successful finite-resistance attack against the system.
The position that frequencies below $f_c$ do actually propagate is also
supported by the fact that, at low frequencies, a coaxial cable is known
to only support TEM modes---these modes are known to have no low frequency
cutoff~\cite[p.~358]{jackson-classical-electrodynamics}.  An exception
occurs when the two ends of the line are held at equal
potential; only standing waves possessing a frequency
that is an integer multiple of ${\nu}/({2L})$ can fulfill these boundary
conditions~\cite[p.~31]{griffiths-quantum-mechanics}.  However, the 
the KKD system differs in allowing arbitrary potentials to appear at the ends
of the line, and so does not support standing waves at the frequency of
operation.

Several attacks against the KKD system exist, however none thus far have
been shown experimentally to substantially reduce the security of the
system~\cite{mingesz-model-line}.

The first attacks, proposed by Scheuer and Yariv~\cite{scheuer-how-secure},
rely upon imperfections in the line connecting the two terminals; the
first exploits transients generated by the resistor-switching operation,
while the second exploits the line's finite resistance.  The former
is foiled by the addition of low-pass filters to the
terminals~\cite{kish-response-scheuer-yariv}, while the latter was
shown to leak less than \SI{1}{\percent} of
bits~\cite{kish-response-scheuer-yariv,mingesz-model-line} in
a practical system.

An attack by Hao~\cite{hao-kish-insecure,kish-response-hao}
instead focuses upon imperfections of the terminals; 
inaccuracies in the noise temperatures of Alice and Bob
create an information leak.  However, it was
demonstrated~\cite{kish-response-hao,mingesz-model-line} that
noise can be digitally generated with a sufficiently
accurate effective noise temperature to prevent this attack from being useful in
practice.

A theoretical argument has been made by Bennett and Riedel~\cite{bennett-kljn}
that no purely classical electromagnetic system can be unconditionally
secure due to the structure of Maxwell's equations.  It is argued that
the upper bound on secrecy rate by Maurer~\cite{maurer-key-agreement} must be
zero because of the locally-causal nature of classical electromagnetics,
and so an eavesdropper can perfectly reconstruct the key with the aid of
a directional coupler.
Kish, et al.~\cite{kish-bennett-riedel} responded that a nonzero
secrecy rate is unnecessary in practice, provided it can be achieved
in the ideal limit.

We begin our attack by analyzing the system in Figure~\ref{fig:kljn-sparam} to
determine the forward- and reverse-travelling waves through the transmission line.
Let us denote the equivalent noise voltages of Alice and Bob $V_a(t)$ and $V_b(t)$ respectively,
and the waves injected onto the line $V'_a(t)$ and $V'_b(t)$.  These are related by
\begin{align}
	V'_a(t) &= \frac{1}{2}(1-\Gamma_A) V_a(t) \\
	V'_b(t) &= \frac{1}{2}(1-\Gamma_B) V_b(t) .
\end{align}
Noting that the mean-squared thermal noise voltage is $\langle V^2 \rangle = 4kTBR$,
we find that
\begin{align}
	\left\langle {V'_a}^2 \right\rangle &= kTBZ_0(1-\Gamma_A^2) \\
	\left\langle {V'_b}^2 \right\rangle &= kTBZ_0(1-\Gamma_B^2) .
\end{align}
As the transmission line in the KKD system is short---and so the forward- and
reverse-travelling waves are equal throughout the line except for a loss factor $\alpha$---we may write
the left- and right-travelling waves at Bob's and Alice's ends of the line respectively as
\begin{align}
	V_+(t) &= V'_a(t) + \alpha \Gamma_A V_-(t) \\
	V_-(t) &= V'_b(t) + \alpha \Gamma_B V_+(t) \\
\intertext{and so}
	V_+(t) &= \frac{V'_a(t) + \alpha \Gamma_A V'_b(t)}{1-\alpha^2 \Gamma_A \Gamma_B} \\
	V_-(t) &= \frac{V'_b(t) + \alpha \Gamma_B V'_a(t)}{1-\alpha^2 \Gamma_A \Gamma_B} \label{eqn:voltages} .
\end{align}
We may write this in matrix form ${\bf v}_d(t) = A{\bf v}_i(t)$ and so find the covariance matrix
$\mathcal{C} = A \mathcal{C}_i A^t$ of the directional components:
\begin{widetext}
\begin{align}
	\mathcal{C} &= \frac{kTBZ_0}{\left( 1-\alpha^2 \Gamma_A\Gamma_B \right)^2} \left[\begin{matrix}
1-\alpha^2\Gamma_A^2\Gamma_B^2 + (\alpha^2-1)\Gamma_A^2 & \alpha \Gamma_A (1-\Gamma_B^2) + \alpha \Gamma_B (1-\Gamma_A^2) \\
\alpha \Gamma_A (1-\Gamma_B^2) + \alpha \Gamma_B (1-\Gamma_A^2) & 1-\alpha^2\Gamma_A^2\Gamma_B^2 + (\alpha^2-1)\Gamma_B^2
\end{matrix}\right]. \label{eqn:covariance}
\end{align}
\end{widetext}
When the line is lossless and so  $\alpha=1$, Eqn.~\ref{eqn:covariance} is invariant under permutation
of $\Gamma_A$ and $\Gamma_B$, and so the covariance matrix provides no
information on the choice of resistors.
However, when $\alpha < 1$ this property fails to hold, allowing the choices
of $\Gamma_A$ and $\Gamma_B$ to be determined from the distribution of
$(v_+,v_-)$.

\begin{figure}[h!]
	\centering
	\includegraphics[width=0.55\linewidth]{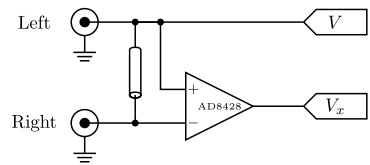}
	\caption{The analog frontend of the directional wave measurement device.  Buffering, offset, gain control,
			and clamping are not shown.  An instrumentation amplifier is used to measure
			the voltage across a \SI{1.5}{\meter} length of coaxial cable,
			providing an estimate of $\partial v/\partial x$.  After offset and gain
			adjustments, the signals are simultaneously sampled 
			by the 12-bit ADCs of an STM32F407 microcontroller.}
	\label{fig:analogue}
\end{figure}

A directional coupler separates forward- and reverse-travelling
waves on a transmission line~\cite{pozar}.  We have constructed
a similar device using differential measurements across a delay line, shown
in Figure~\ref{fig:analogue}.

Consider the d'Alembert solution~\cite[Eqn.~7.7]{jackson-classical-electrodynamics} to the wave equation in a
medium with propagation velocity $\nu$,
\begin{align}
	v(t,x) &= v_+\left(t - \frac{x}{\nu}\right) + v_-\left(t + \frac{x}{\nu}\right) .
\end{align}
The forward-travelling component $v_+(\tau)$ differs from the reverse-travelling
component $v_-(\tau)$ in the sign of its spatial argument.  We use this to our
advantage by computing the linear combinations
\begin{align}
	\frac{\partial v}{\partial t} - \nu\frac{\partial v}{\partial x} &= 2\frac{d v_+}{d t} \\
	\frac{\partial v}{\partial t} + \nu\frac{\partial v}{\partial x} &= 2\frac{d v_-}{d t},
\end{align}
yielding the forward- and reverse-travelling waves as we desire.  All that
remains, then, is to determine $\partial v/\partial t$ and $\partial v/\partial x$.

The time derivative $\partial v/\partial t$ may be determined digitally from
sampled values of $v(t)$.  The spatial derivative is approximated as being
proportional to the voltage across a short delay line, shown in
Figure~\ref{fig:analogue}.

After digitisation, we high-pass filter the signals $V$ and $V_x$
in order to remove any DC offsets or mains interference.
The signals are then combined to produce the left- and
right-travelling waves.  The time-derivative $\partial v/\partial t$ can be
approximated by a difference operator, however in order to accommodate for
the unknown propagation velocity and delay line length, common-mode leakage
into $V_x$, and losses in the delay line, we instead use a first-order
least-mean-squares (LMS) adaptive filter~\cite{haykin-adaptive-filter}
for initial calibration.  A signal source is applied to one port
and the other is terminated; this produces a right-travelling wave
on the line, but none travelling to the left.  The left-travelling
output $V_-$ is used as an error signal for the LMS filter,
suppressing any contribution from the right-travelling wave.

\begin{figure}[h!]
	\centering

	\includegraphics[width=0.7\linewidth]{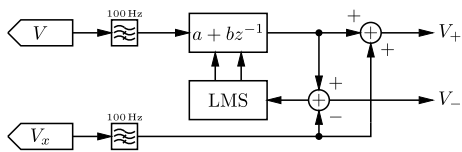}
	\caption{The digital signal processing of the directional wave measurement device, implemented
			on an STM32F407 microcontroller.  Offset
			removal is not shown.  A least-mean-squares filter is used
			at startup to determine the necessary filter coefficients;
			a signal is applied to one port while the other is connected
			to a terminator, and the filter coefficients adjusted to force
			$V_- = 0$.  Filter updates are disabled once the apparent
			reflection coefficient becomes sufficiently small.}
	\label{fig:dsp-block}
\end{figure}

The real part of the reflection coefficient,
seen looking out of the right port,
is computed by a cross-correlation
between left- and right-travelling waves.  When this
falls below $0.01$, calibration is declared
complete and filter updates cease.  After calibration, we validate the system by configuring it as
a reflectometer.  Open and shorted measurements
are made, yielding reflection coefficients of $+1$ and $-1$
respectively.  The reflection coefficients of several resistors are
also measured, again yielding the expected values.

We have described the implementation of a directional
wave measurement device using differential measurements across a delay line.
While we might measure the power travelling in each direction in
order to determine the resistor configuration, the
distributions to be distinguished are very similar,
resulting in a relatively large bit-error rate (BER)
as was shown in~\cite{kish-response-scheuer-yariv}.  However, 
comparison of the variances of $v_+$ and $v_-$ is suboptimal.
We derive an improved test using Bayesian methods and demonstrate
that the two cases can be far more easily distinguished.

Knowing the covariance matrices of $v_+(t)$ and $v_-(t)$ for each hypothesis,
we may use Bayes' theorem~\cite{larsen-mathematical-statistics} to determine
the probability of each configuration.  Let $C=0$ and $C=1$ refer
to the events that $(R_a,R_b) = (R_h,R_l)$ and vice-versa, respectively.
Then,
\begin{align}
	P[C=0 | {\bf v_+} \cap {\bf v_-}] 		&= \frac{P[{\bf v_+}\cap {\bf v_-} | C=0]P[C=0]}{P[{\bf v_+}\cap {\bf v_-}]} \\
		&= \frac{\frac{1}{2} p_0({\bf v_+},{\bf v_-})}{\frac{1}{2} p_0({\bf v_+},{\bf v_-}) + \frac{1}{2} p_1({\bf v_+},{\bf v_-})} \\
		&= \frac{1}{1 + \frac{p_1({\bf v_+},{\bf v_-})}{p_0({\bf v_+},{\bf v_-})} }\label{eqn:bayesian} ,
\end{align}
where $p_0(\cdot,\cdot)$ and $p_1(\cdot,\cdot)$ are the multivariate Gaussian
PDFs for the measurements from each respective configuration.

The most probable state, then, is given by the maximum-likelihood
estimator~\cite{larsen-mathematical-statistics}
\begin{align}
	\hat{C} =
		\begin{cases}
			0 & \text{if } p_0({\bf v_+},{\bf v_-}) > p_1({\bf v_+},{\bf v_-}) \\
			1 & \text{if } p_0({\bf v_+},{\bf v_-}) < p_1({\bf v_+},{\bf v_-}) . \\
		\end{cases} \label{eqn:estimator}
\end{align}
The comparison is more conveniently made in terms of the log-likelihood,
which for the $n$-variate zero-mean Gaussian distribution with
covariance matrix $\Sigma$ is given by~\cite[p.~250]{cover-information-theory}
\begin{align}
    \log p_\Sigma({\bf x})
        &= \log\left[ \frac{1}{\left(2\pi\right)^\frac{n}{2} \left|K\right|^\frac{1}{2}}
                e^{-\frac{1}{2}{\bf x}^T \Sigma^{-1} {\bf x}} \right] \\
        &= -\frac{1}{2} \log \left|\Sigma\right| -\frac{n}{2} \log \left(2\pi\right)
            -\frac{1}{2} {\bf x}^T \Sigma^{-1} {\bf x} .
\intertext{Noting that $\Sigma$ is positive-definite, we may write it
            in terms of its Cholesky decomposition $\Sigma = K K^T$, and so}
        &= -\frac{1}{2} \log \left|\Sigma\right| -\frac{n}{2} \log \left(2\pi\right)
            -\frac{1}{2} \left\Vert K^{-1} {\bf x} \right\Vert^2 . \label{eqn:ll-power}
\end{align}
Only the final term depends upon the data, and there only through the total
power of a group of signals $K^{-1} {\bf x}$ formed by linear combinations of
the measured waves.

\begin{figure}
    \centering
    \subfloat[Uncorrelated measurements]{
        \includegraphics[width=0.7\linewidth]{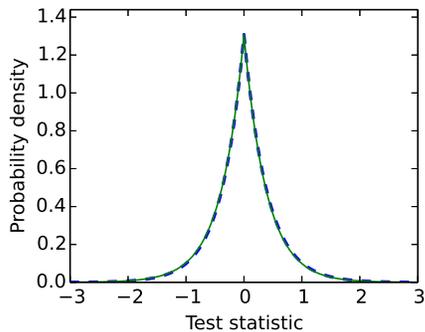}
        \label{fig:estimator:uncorrelated}
    }
    
    \subfloat[Correlated measurements]{
        \includegraphics[width=0.7\linewidth]{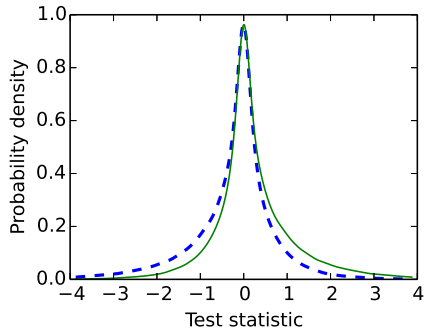}
        \label{fig:estimator:correlated}
    }
    
    \caption{Log likelihood-ratio test statistics for each permutation
            of resistors in Eqn.~\ref{eqn:covariance}, as in
            Eqn.~\ref{eqn:ll-power} with scaling-factors omitted.
            The dashed lines correspond to the case where
            $(R_a,R_b) = (R_l,R_h)$, and the solid lines to
            $(R_a,R_b) = (R_h,R_l)$.
            Parameters are $R_l = \SI{1}{\kilo\ohm},
            R_h = \SI{10}{\kilo\ohm}, Z_0 = \SI{50}{\ohm}$,
            and $\alpha = \SI{-0.1}{\decibel}$.
            In Figure~\ref{fig:estimator:uncorrelated} the
            covariances are set to zero, and so
            Eqn.~\ref{eqn:estimator} reduces to a simple power
            comparison.  The distributions are almost
            indistinguishable.  In Figure~\ref{fig:estimator:correlated},
            the measurement variables are drawn from a correlated
            bivariate distribution having the same marginal
            variances, and are far more distinguishable.  In either
            case, as losses increase and so the variances of the measurements
            and transformed measurements respectively differ more greatly,
            the two distributions, which mirror each other about zero, 
            become increasingly assymmetric and so far more distinguishable.}
    \label{fig:estimator}
\end{figure}

It should be noted that this estimator differs substantially from that
proposed in~\cite{scheuer-how-secure}, which makes a simple comparison of
variances.  The measured variables in our case are collected simultaneously
and so exhibit the heavy correlations of Eqn.~\ref{eqn:covariance}.
With these correlations, the likelihood-ratio test provides far better performance
than the difference in the variances of the marginal distributions would
suggest. However, if the voltage and current measurements are considered
separately, as in~\cite{kish-response-scheuer-yariv,mingesz-model-line}
where only the marginal distributions of each measurement
are computed, these correlations vanish and
so the estimator described in Eqns.~\ref{eqn:estimator} and~\ref{eqn:ll-power}
has substantially less power.
The distribution of test statistics is shown in Figure~\ref{fig:estimator}
for a loss of \SI{0.1}{\decibel}.
The presence of correlation causes the distributions of test statistics
to differ substantially, where otherwise they would be almost
indistinguishable.

\begin{figure}
	\centering
	\includegraphics[width=0.7\linewidth]{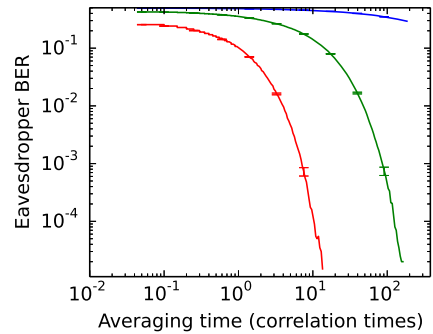}
	\caption{Simulated eavesdropper bit-error-rate as a function of averaging time, for
			line attenuations of 0.01, 0.1, and 1.0
			decibels respectively from top to bottom.  The link parameters
            are $R_L = \SI{1}{\kilo\ohm}, R_H = \SI{1}{\kilo\ohm}, Z_0 = \SI{50}{\ohm}$.
            Note that the averaging
			time is expressed in multiples of \SI{200}{\micro\second}.
			This is the correlation time (i.e.~reciprocal of the system bandwidth) so that the results are bandwidth independent.
			Transmission lines with greater loss are more susceptible to attack,
			with substantial attenuations providing little protection.  The
            error rates are estimated from a sample size of $10^5$, with
            $2\sigma$ error bars shown.}
	\label{fig:ber-vs-loss}
\end{figure}

The results of simulation for various
values of loss are shown in Figure~\ref{fig:ber-vs-loss}.
A pair of white noise processes are generated, Fourier-transformed,
and the undesirable frequency components removed.  They are combined according to Eqn.~\ref{eqn:voltages}
to produce the voltage waves,
and the maximum-likelihood estimator is used to determine the resistor
configurations.  This demonstrates that our estimator can
differentiate the two distributions without the unreasonably large
sample sizes that were previously thought necessary~\cite{kish-response-scheuer-yariv}.

\begin{figure}
	\centering
	\includegraphics[width=0.7\linewidth]{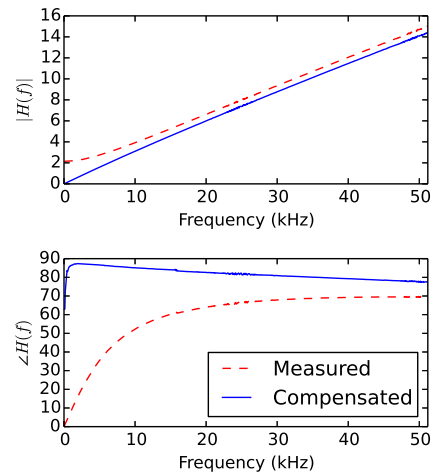}
	\caption{Measured frequency response of the $\partial v/\partial x$
			estimation circuit in Figure~\ref{fig:analogue}.
			The derivative increases linearly with
			frequency, as would be expected from the
			d'Alembert solution to the wave equation.
			The response $H(0)$ at DC is subtracted in order
			to remove the effect of wire resistance, yielding
			the `compensated' curves above.  After this
			correction we see $\angle H(f)$
			approximating the expected $+90^\circ$ constant
			phase response, slightly drooping due to the limited
            frequency response of the system.}
	\label{fig:vx-response}
\end{figure}

Having demonstrated our attack in simulation, we proceed to
experimental validation of the model.
The estimation of $\partial v/\partial x$ is key to the operation of the device,
however the synthesis provided above is dependent upon a wave-based
analysis of the system.  We therefore measure experimentally the frequency response of
the electronically-estimated $\partial v/\partial x$,
shown in Figure~\ref{fig:vx-response}, with a wave travelling in a single direction
in order to verify that our analysis is appropriate.

We expect to see a magnitude response linear in frequency
and a constant $+90^\circ$ phase response.  This agrees with
the experimental results shown in Figure~\ref{fig:vx-response},
validating our analysis, and demonstrates that the signal
through a short transmission line indeed propagates as a wave,
in contradiction to the theoretical claims of~\cite{kish-notes-on-recent-approaches}.

\begin{figure}
	\centering
	\includegraphics[width=0.7\linewidth]{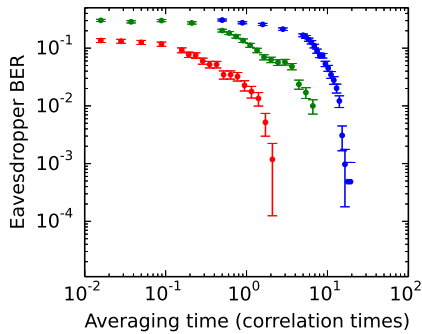}
	\caption{Measured eavesdropper bit-error-rate as a function of averaging time
			and line attenuation.
			The line is approximately \SI{2}{\meter} in length
			and has a loss of less than \SI{0.1}{\decibel}.
			From top to bottom, \SI{0}{\decibel}, \SI{0.1}{\decibel},
			and \SI{1}{\decibel} of additional attenuation provided
            by inserting an in-line attenuator at one end of the
            line.}
	\label{fig:eavesdropper-ber-experimental}
\end{figure}

We have implemented the attack described above, using resistances
$R_l = \SI{1}{\kilo\ohm}$, $R_h = \SI{10}{\kilo\ohm}$, and
a coaxial transmission line of characteristic impedance $Z_0 = \SI{50}{\ohm}$.
The voltage sources are produced by an arbitrary waveform generator,
producing independent normally-distributed voltages over a frequency
range of \SI{500}{\hertz}--\SI{5500}{\hertz}.  The bandwidth
$B=\SI{5}{\kilo\hertz}$ results in an approximate correlation time
of $B^{-1} = \SI{200}{\micro\second}$~\cite{kish-kljn-mitm}.
Each configuration is set and the
covariance matrices from Eqn.~\ref{eqn:covariance} are measured during
the setup phase.
Resistor configurations are randomly selected as
would be the case in an operational system, and the log-likelihood ratios
are computed for the measured values of $v_+$ and $v_-$.  Their differences
are thresholded to compute~(\ref{eqn:estimator}), providing the bit-error rates in
Figure~\ref{fig:eavesdropper-ber-experimental}.  Even modest
losses allowed almost all bits to be determined correctly,
showing that the technique simulated in Figure~\ref{fig:ber-vs-loss}
can be applied in practice.

By applying a threshold to the likelihood ratios, we may estimate the agreed
bits and so determine the error rate of Eve.  We see that even with
the minuscule losses of the test system,
Eve can acquire a substantial proportion of the agreed bits.

The technique above exploits imperfections in the KKD implementation;
while it might be theoretically possible to counter this attack
by reduction of losses as proposed in~\cite{kish-response-scheuer-yariv}, the 
reduction of losses substantially below \SI{0.1}{\decibel} ensures that this
will be infeasible for all but the shortest or slowest of links.

This raises the question of why our attack should succeed where existing
finite-resistance attacks have failed.  The attack of Scheuer and
Yariv~\cite{scheuer-how-secure} considered only the variances of
the measured variables.  Our attack exploits the large correlation
between waves in each direction; the estimator used above
partially removes this common signal, increasing
the ability to distinguish between the
two cases statistically.

We have demonstrated an attack against the KKD key distribution system
that exploits losses within the connecting transmission
line.  The attack has been shown experimentally to correctly
determine more than \SI{99.9}{\percent} of bits transmitted over a
\SI{2}{\meter} transmission line within 20 correlation times.
As this attack requires that losses be reduced to a fraction of
a decibel in order to maintain a meaningful level of
security, modifications to the system will be necessary in order to
produce a secure link of any significant length and bitrate.


%

\end{document}